\documentstyle[prl,twocolumn,aps,tighten,floats,epsfig,psfig]{revtex}
\begin{document}
\draft 
\title{Designing emissive conjugated polymers with small
optical gaps: a step towards organic polymeric infrared lasers}

\author{Alok Shukla$^{1,2}$ and Sumit Mazumdar$^{1}$}
\address{$^1$Department of Physics and The Optical Sciences Center, 
University of Arizona, Tucson, AZ 85721}
\address{$^2$Cooperative Excitation Project ERATO, Japan Science and
Technology Corporation (JST)} 

\maketitle

\begin{abstract}
We show that chemical modification of the trans-polyacetylene structure
that involves substitution of the backbone hydrogen atoms with conjugated
side groups, leads to reduction of the backbone bond alternation as well
as screening of the effective Coulomb interaction. Consequently
the optical gap of the substituted material is smaller than the parent
polyene with the same backbone length, and the excited state ordering
is conducive to efficient photoluminescence. The design of
organic polymeric infrared lasers, in the ideal long chain limit,
thereby becomes possible.
\end{abstract}
Much of the current experimental work on conjugated polymers
centers around the
strong photoluminescence (PL) observed in specific materials. A
complete physical understanding of the structural characteristics necessary
for luminescence is clearly desirable for designing materials with emission
wavelengths that can vary over a wide range. 
Within existing theories,
light emission
should be restricted to systems with optical gaps
larger than that of the nonemissive polymer trans-polyacetylene (t-PA). 
In the
present Letter we show that suitable
molecular engineering 
can 
screen the many-body Coulomb interactions,
thereby lifting this
restriction.

The distinction between nonluminescent and luminescent conjugated
polymers is related to the Coulomb correlations between
the $\pi$-electrons in these systems. 
Excited states of these centrosymmetric systems are either one-photon allowed
and of B$_u$ symmetry, or two-photon allowed and of A$_g$ symmetry.
A consequence of moderate
Coulomb interactions is that the 
the 2A$_g$ excited state occurs below the 1B$_u$ state in linear polyenes and
t-PA \cite{Kohler1,Schulten1,Ramasesha}. The optically pumped 1B$_u$ state here
rapidly decays to the 2A$_g$ instead of to the ground state. Radiative
transition from the 2A$_g$ to the 1A$_g$ ground state is forbidden, and
polyenes and t-PA are therefore
nonemissive. Strong PL in the common light emitting polymers, such as the
polyparaphenylenes (PPP) and
poly(para-phenylenevinylenes) (PPV), implies a reversed excited state
ordering E(2A$_g$) $>$ E(1B$_u$) (where E(...) is the energy of the state),
which is understood within the effective
linear chain models for these systems \cite{Soos2}. Within this theory, the
lowest excitations of the above class of materials can be mapped into
those of polyene-like chains with large effective bond alternations.
For example, replacement of all (alternate) double
bonds of polyacetylene with phenyl groups gives PPP (PPV), a chemical
modification that we hereafter refer to as ``bond substitution''.
Excited state orderings of the effective linear chain are understood 
most simply
within the
dimerized Hubbard model
\begin{equation}
H = - \sum_{i} t(1 \pm \delta)(c_{i,\sigma}^{\dag}c_{i+1,\sigma}
+ c_{i+1,\sigma}^{\dag}c_{i,\sigma}) + U \sum_i n_{i,\uparrow} n_{i,\downarrow}
\label{eq-hub}
\end{equation}
where $c_{i,\sigma}^{\dag}$ creates a $\pi$ electron of spin $\sigma$ on
carbon atom $i$, $n_{i,\sigma}=c_{i,\sigma}^{\dag}c_{i,\sigma}$, $t$ is the
nearest neighbor hopping matrix element, and
$\delta$ is the effective bond alternation parameter. 
For small $\delta$, the 2A$_g$
is a spin excitation that occurs below the charge excitation 1B$_u$
\cite{Schulten1,Ramasesha}. For
large $\delta$, the 2A$_g$ acquires charge excitation character, and 
occurs above the 1B$_u$ \cite{Soos2,Shuai}.

The prescription for obtaining light emission within existing theories is
then to increase the effective $\delta$.
Since larger
$\delta$ increases E(1B$_u$), this prescription restricts light
emission from polymers whose optical gaps 
are necessarily larger than that of t-PA. Whether or not it
should be possible to design light emitting materials with {\it smaller}
optical gaps, at least theoretically, is clearly an intriguing question.
The necessary conditions for this are, (i)
$\delta$ same or even smaller than in linear polyenes, and (ii) an
{\it effective} Hubbard correlation parameter $U_{eff}$ that is also smaller
than in the polyenes, such that E(2A$_g$) $>$ E(1B$_u$), in spite of
smaller $\delta$. In the present Letter,
we show how both of these can be achieved through ``site-substitution'', in
which individual CH groups of the parent polyene structure (as opposed to
pairs of them) are modified.

\begin{figure}
\psfig{file=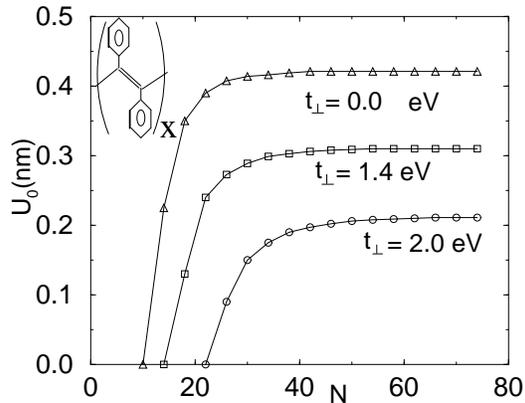,width=7.0cm,angle=0}
\vskip 10pt
\caption{Dimerization amplitude $u_0$ for
PDPA periodic rings with N carbon atoms on the polyene backbone,
for different $t_{\perp}$ 
The inset shows the chemical structure of 
PDPA.}
\label{fig-ssh} 
\end{figure}

Although our work is general, for clear illustration
of the principle, we focus here on the recently synthesized 
poly(diphenylacetylene) (PDPA)
(see inset Fig. \ref{fig-ssh}) class of conjugated polymers
\cite{Tada}. The current materials have short conjugation lengths 
(five to seven backbone double bonds \cite{Tada}), but
the observation of PL
quantum efficiency $>$ 50\% in several PDPA oligomers \cite{Tada,Valy}
is nevertheless counterintuitive, given their polyene backbones. 
The estimated
conjugation lengths 
are large enough that in the corresponding polyenes
E(1B$_u$) $>$ E(2A$_g$) \cite{Kohler2}. Since the Hubbard U is an {\it atomic}
property, naively one would have expected the energy ordering in PDPA to be
the same as in the corresponding polyene, and therefore weak PL. 
An alternate
naive interpretation, viz., the relative location of the 2A$_g$ in PDPA
will be somewhere
in between that of trans-stilbene (the unit cell of PDPA) and linear polyenes,
is incorrect, since 
this interpretation implies that the absolute 1B$_u$ energy in PDPA is
higher than that in the polyene
with the same chain length and lower than that in trans-stilbene.
However, both experimentally
\cite{Tada,Valy} and theoretically (see below), the 1B$_u$ in PDPA occurs
{\it below} that of the corresponding polyene.
We show here that the correct interpretation involves a reduction in the
effective Coulomb correlation.

We consider the Pariser-Parr-Pople
Hamiltonian for PDPA, supplemented by the Su-Schrieffer-Heeger (SSH) 
electron-phonon (e-ph) coupling \cite{SSH}
\begin{mathletters}
\label{allequations}
\begin{equation}
H = H_1 + H_2 + H_3  + H_{ee},  \label{eq-ham}
\end{equation}
\begin{equation}
H_1 = -\sum_{\langle k,k' \rangle,M} (t_0 - \alpha \Delta_{k,M}) 
B_{k,k';M,M+1}
 + \frac{1}{2}  K  \sum_{k,M} \Delta_{k,M}^2, \label{eq-h1}
\end{equation}
\begin{equation}
H_2=-t_0 \sum_{\langle\mu,\nu\rangle,M} B_{\mu,\nu;M,M}, \label{eq-h2}
\end{equation}
\begin{equation}
H_3= -t_{\perp} \sum_{M} B_{k,\mu,M,M,}, \label{eq-h3}
\end{equation}
\begin{equation}
H_{ee}=U \sum_{i,M} n_{i,M,\uparrow} n_{i,M,\downarrow}
+ \frac{1}{2} \sum_{i \neq j,M,N} V_{i,j,M,N}(n_{i,M}-1)(n_{j,N}-1), \label{eq-hee}
\end{equation}
\end{mathletters}
In the above, $k$, $k'$ are carbon atoms on the polyene backbone, 
$\mu, \nu$ are carbon atoms belonging to the phenyl groups, $M$ is a
unit 
consisting of a phenyl group and a polyene carbon, $\langle ... \rangle$
implies nearest neighbors, and 
$B_{i,j,M,M'} = \sum_{\sigma}(c_{i,M,\sigma}^\dagger c_{j,M',\sigma} + 
h.c.)$.
$H_1$ and $H_2$ describe the electron hoppings
within the polyene backbone and the phenyl rings, respectively, and $H_3$ is
the hopping between them. In $H_1$,  
$\alpha$ is the e-ph
coupling constant, $K$ the spring constant, and 
$\Delta_{k,M}=(u_{k+1,M+1}-u_{k,M})$, where $u_{k,M}$ is the displacement
of the chain carbon atom of the Mth unit from equilibrium.
$H_{ee}$ is the e-e interaction, with $i$ and $M$ including now all
atoms. Steric interactions 
imply considerable phenyl group rotation in PDPA. Within 
Eq.~(2) this affects only $V_{i,j,M,N}$ 
and $t_{\perp}$. 
In the following we first show that bond alternation in PDPA is smaller than in 
t-PA, thereby proving that the observed PL is not due to enhanced bond
alternation.
We then present a single configuration interaction (SCI) theory of the
1B$_u$ that provides the mechanism of the reduction of $U_{eff}$. Finally, we 
present many-body calculations of excited state orderings, and of
a pair correlation function that directly demonstrates reduced $U_{eff}$ in
PDPA.
 
We compare the
bond-alternations in t-PA and PDPA 
for $H_{ee}$ = 0 within Eq.~(2) 
and postpone the justification of ignoring $H_{ee}$.
We have chosen $t_0$ = 2.4 eV,
$\alpha$ = 4.1 eV/A and
$K$ = 21 eV/A$^2$. 
It is difficult to estimate $t_{\perp}$, but it is smaller than
the usual ``single''
bond. In Fig. \ref{fig-ssh} we show our calculated bond alternations $u_0$
[$u_i$ = (--1)$^iu_0$ in Eq.~(2)] for perfectly dimerized PDPA
periodic rings with
chain lengths of N backbone carbon atoms, 
for $t_{\perp}$ = 0, 1.4 eV and 2.0 eV. The calculated $u_0$ decreases
rapidly with
increasing $t_{\perp}$.

An {\it a posteriori} justification of the neglect of $H_{ee}$
in the above calculations can now be given. 
In previous study of Coulomb effects on bond alternation in the infinite
polyene \cite{Mazumdar2}, it was shown
that for all parametrizations of the Pariser-Parr-Pople Hamiltonian
bond alternation is enhanced compared to the SSH prediction.
More importantly, these parameters
belong in a regime where 
further increasing of Coulomb
interactions enhances the bond alternation even more. 
Since we show below
that the effective correlations in PDPA are {\it smaller} than that in t-PA, 
inclusion of 
$H_{ee}$ would lead to even smaller bond alternation in PDPA than in t-PA.

A preliminary understanding of 
the many-body effects in 
PDPA can be obtained
by a detailed examination of the 1B$_u$ state alone.
We calculate the transition dipole couplings between the ground state
and the 1B$_u$, for the five double-bond oligomer.
Our calculations
are for both zero and nonzero $H_{ee}$. We chose rigid bond alternation
$\alpha u_0/t_0$ = 
$\delta$ = 0.07, the same as in linear polyenes.
The Coulomb 
parameters were chosen from the Ohno relationship \cite{Ohno},
\begin{equation}
V_{i,j,M,N} = U/(1+0.6117R_{i,j,M,N}^2)^{1/2} \; \mbox{,}
\label{eq-ohno}
\end{equation}
where $U$ = 11.13 eV and $R_{i,j,M,N}$ is the separation in A. between
carbon atoms $i$ on unit M and $j$ on unit N.
The backbone polyene bond lengths are assumed to be 1.45 and 1.35 A. in
length, as in t-PA, and all phenyl bond lengths are taken to be 1.4 A. The
bond between the backbone atoms and the phenyl groups is taken to be a
true single bond with length 1.54 A. We considered both parallel and
antiparallel ordered rotations of neighboring phenyl groups, with rotation
angles of 30$^o$. The difference in the dipole moments for these two cases
was insignificant. 
The nonzero $H_{ee}$ calculations were done within the
single configuration interaction (SCI) approximation, which is known to give
a reasonable description of the 1B$_u$. We define the backbone axis as the
x-axis and the transverse direction as the y-axis.  
The results of our calculations of transition dipole couplings $\mu_x$ and
$\mu_y$ along x- and y-directions
are summarized in
Table \ref{tab-edm} for several $t_{\perp}$. 
We have included in Table \ref{tab-edm} the
calculated results for an artificial
trans-stilbene monomer in which the hopping integral between the ethylenic
linkage and the phenyl groups is 1.4 eV (the y-axis corresponds to the long
molecular axis of trans-stilbene).

The PDPA transition dipole couplings in Table \ref{tab-edm} have
significant y-contribution. 
For all 
$t_{\perp}$, 
$H_{ee}$ strongly decreases $\mu_x$, 
a well-known result in the case of linear polyenes: the transition 
dipole moment is a direct measure of the particle-hole separation in the
1B$_u$, and
since $H_{ee}$ leads to particle-hole confinement and
exciton formation, the reduction in $\mu_x$
is expected.
More interesting conclusions emerge 
upon 
comparisons of the different $t_{\perp}$ cases.
For nonzero $H_{ee}$, the particle-hole confinement in the 
x-direction is 
significantly {\it enhanced} by $t_{\perp}$, as indicated by the 
strong reduction
in $\mu_x$ as $t_{\perp}$ is increased from zero. 
Simultaneously,
there is an effective ``deconfinement'' in the y-direction, as evidenced
by the increase in $\mu_y$. This deconfinement
is large enough that $\mu_y$ is {\it larger} for the interacting Hamiltonian 
than for the noninteracting model for all nonzero $t_{\perp}$,
indicating that the confinement along the x-axis and the deconfinement along
the y-axis are {\it synergic}.
The synergic nature of the deconfinement is
further confirmed upon comparison with 
the transition dipole
moments of 
the trans-stilbene monomer: 
$H_{ee}$ suppresses $\mu_y$ in the isolated
monomer, but enhances it in PDPA. 
Physically, as the exciton
is ``squeezed'' along the x-direction, it spreads out in the y-direction.

The deconfinement along the transverse direction is a signature
of reduced Coulomb effects in PDPA.
In any mapping of the excited states of PDPA to that
of an effective linear chain, a {\it composite} site is to be constructed 
from each carbon atom on the polyene backbone and the carbon atoms of the
phenyl group.
For moderate $U$ the particle and the hole correspond to a
double occupancy and a vacancy in the 1/2-filled band, which in PDPA are
delocalized over the phenyl segments. The 
synergic deconfinement then indicates that (i) the on-site repulsion
$U_{eff}$ on a composite site of an effective linear chain 
is of the form U -- W$_{\perp}$, and (ii)
$W_{\perp}$ is large.

We confirm this conjecture by direct evaluations of E(2A$_g$) and E(1B$_u$).
Electron correlation calculations considerably more sophisticated than the
SCI are required for the determination of the energy of the highly correlated 
2A$_g$ state. Such calculations are not possible for the PDPA oligomer
with five double bonds. We therefore compare the PDPA oligomer with 
four backbone carbons and butadiene within the rigid band 
approximation for Eq.~(2). 
Such a calculation is representative, since any reduction in $U_{eff}$ 
occurs at every composite site of longer oligomers. Even for this
short oligomer with twenty eight carbon atoms, however,  
it is not possible to perform 
full-CI (FCI) calculations retaining all the orbitals. Instead of making
further questionable approximations 
we  perform three different
FCI calculations, retaining only a limited set of molecular
orbitals (MOs) in each case. 
The three calculations, taken together, provide an unambiguous picture.

\begin{figure}
\psfig{file=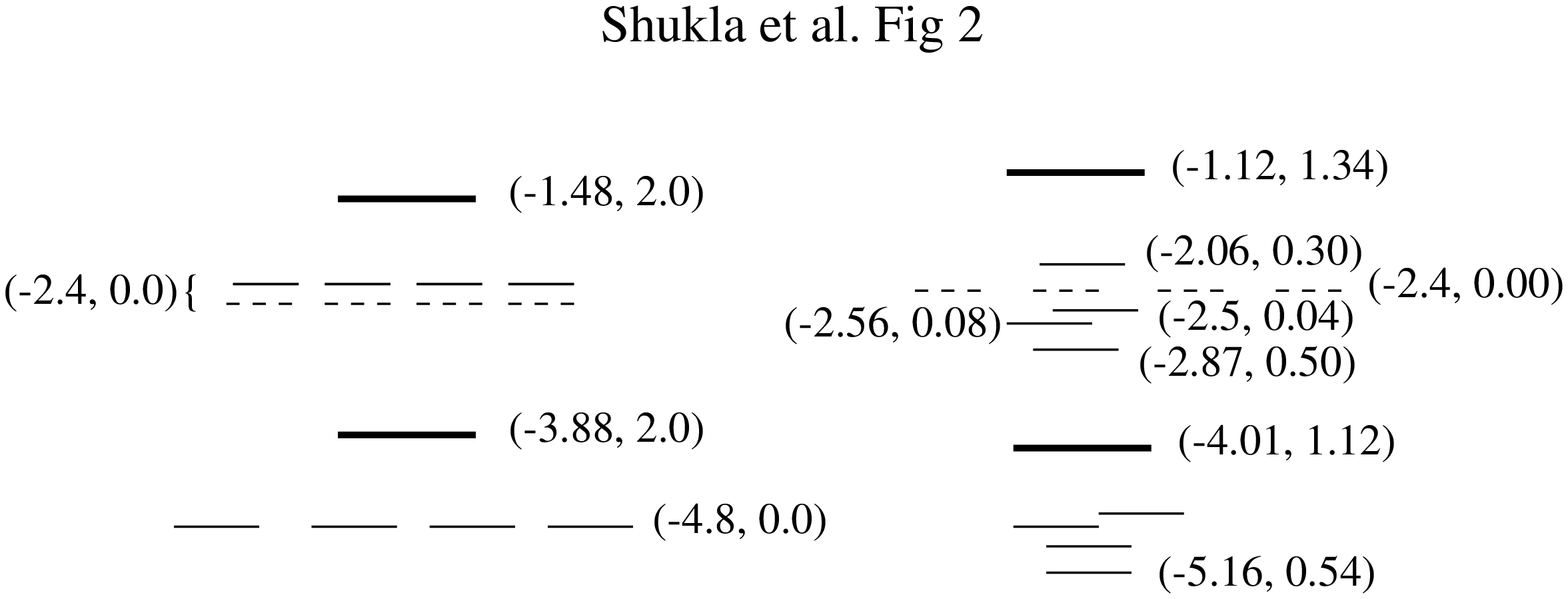,width=3.8cm,angle=-90}
\caption{Bonding MOs of a PDPA oligomer with four backbone carbon atoms,
for $\delta$ =0, and $t_{\perp}$ = 0.0 (left) 1.4 eV (right).
The first number within each parenthesis is the one-electron energy in eV, 
while the second number is the contribution of the MO to the 
backbone electron density.
Chain-like MOs (thick
solid lines), delocalized benzene-derived MOs (regular solid lines), and
localized benzene MOs (dashed lines) are identified for 
nonzero $t_{\perp}$ from their one-electron energies and contributions to the
backbone electron density.}  
\label{fig-elev} 
\end{figure}

In Figs.~\ref{fig-elev} we have shown the bonding MOs of the 
PDPA oligomer with four backbone carbons, for $\delta$ = 0, and for
$t_{\perp}$ = 0 and 1.4 eV.
For nonzero $t_{\perp}$ there is considerable
mixing between the backbone and phenyl MOs, but nevertheless, it is possible
to identify the two chain-derived MOs from the large separations of their
one-electron energies from those of the benzene-derived MOs,
as well as from their large contributions to the electron 
densities on the backbone carbon 
atoms (1.34 and 1.12). There exist only three other bonding
MOs whose contributions to the backbone carbon electron densities
(0.30, 0.50 and 0.54) are substantial 
Our first set of FCI calculations (hereafter set 1)
involve all ten bonding and antibonding MOs with strong chain components.
A second set 
(set 2) were done with
the two chain-derived bonding MOs and the four highest delocalized 
benzene-derived bonding
MOs (only two of which contribute to chain carbon electron densities),
and the corresponding six antibonding MOs. Set 3
calculations involved the two chain-like MOs and the four localized
benzene MOs,
and the corresponding antibonding MOs. 

The results of our calculations for Ohno parameters 
are shown
in Table \ref{tab-elev1}, where we have included the exact results for
butadiene.
Only for $\delta$ = 0 within set 1 is the 2A$_g$
in PDPA below the 1B$_u$, while for all other cases E(2A$_g$) $>$ E(1B$_u$),
in contrast to butadiene, where E(2A$_g$) is substantially smaller than
E(1B$_u$). We emphasize that the MOs retained within sets 1, 2 and 3, with the
exclusion of the chain-derived MOs, are mutually exclusive. 
Therefore, in the absence of 
peculiar and unexpected interference effects, standard CI concepts predict that
a complete calculation performed
with all the MOs would also
find the 2A$_g$ above the 1B$_u$.
Although it may be argued 
that in the longer PDPA oligomers the relative E(2A$_g$) is smaller, 
we emphasize that the calculations in Table \ref{tab-elev1} are
for a rather small $t_{\perp}$ (1.4 eV), and the realistic $t_{\perp}$ can
be larger, enhancing E(2A$_g$).
The cosine dependence of hopping integrals on the rotational
angle that is often assumed predicts $t_{\perp}$ = 1.9 eV for a rotational
angle of 30$^o$. The trend indicated in Table \ref{tab-elev1} is valid for
all chain lengths and $\delta$.

Finally, we substantiate our argument for a reduced $U_{eff}$ in PDPA
from calculations of the ground state pair correlation function $g =  
(1/N) \sum_i \langle n_{i,\uparrow}n_{i,\downarrow}\rangle$ 
($N$ = total number
of sites)
Larger $g$ implies smaller $U_{eff}$ 
(for the half-filled band, $g$ = 0.25 at $U$ = 0, while $g$ = 0 
for $U \to \infty$) \cite{Dixit}. 
In order to demonstrate
that the reduced $U_{eff}$ is due to deconfinement of the
double occupancy, and not due to the additional intersite Coulomb interactions
in PDPA, these calculations were done 
for $V_{i,j,M,N}$ = 0.
The pair-correlation functions were calculated from the relationship
$g = (1/N) \partial 
\langle H \rangle/\partial U$. 
For $U$ = 2$t_0$, $g$ =0.152 for butadiene and 0.18
for benzene,
while for PDPA this quantity is 0.237 within the set 1
calculation and 0.207 within set 2 (set 3 would give even larger $g$,
from the results of Table \ref{tab-elev1}), suggesting that the exact 
$g$ for PDPA
is in between 0.207 and 0.237. Larger $g$ in PDPA than
that of both the constituents 
is due to the synergic nature of the deconfinement discussed
above. 

Smaller $\delta$ and $U_{eff}$ in PDPA, compared
to t-PA, necessarily implies smaller optical gap. Within 
one-electron theory, and with 
SSH parameters, the optical gap in
PDPA in the long chain limit is 0.9 (0.56) eV for $t_{\perp}$ = 1.4 (2.0)
eV, compared to 1.4 eV for t-PA.
For nonzero $H_{ee}$, we have
calculated E(1B$_u$) within SCI for linear polyenes and PDPA oligomers 
upto eight double bonds, with {\it fixed} $\delta$.
For all chain lengths
the PDPA optical gap is smaller than the polyene with the same number of carbon
atoms. For example, at five (seven) double bonds, our calculated gap for the
polyene is 3.8 (3.4) eV, close to the experimental value \cite{Kohler2}.
The calculated gap in PDPA with five (seven) double bonds 
and $t_{\perp}$ = 1.4 eV is 3.2 (2.9) eV.
With self-consistent (and therefore smaller) $\delta$, the gap in PDPA will get
even smaller. Moreover, within our theory,
poly(1-alkyl,2-phenylacetylene) (PAPA) with conjugated side groups on only
{\it alternate} backbone carbons will have
larger $\delta$ and 
$U_{eff}$. Significantly, optical gaps in PAPA oligomers are uniformly larger 
than those in PDPA \cite{Tada}.
The reduction in $U_{eff}$ found here
will occur in any system modified by
site-substitution (or suitable combination of site and bond-substitution).
While the current PDPA oligomers emit in the visible due to short
conjugation lengths, it is then possible conceptually
to have conjugated polymers emitting in the infrared. The synthetic challenge
would be to obtain 
polyene derivatives which have conjugated side groups but which still possess
long rigid 
segments. 

Work by S.M. was supported by NSF-ECS and the ONR through the MURI center
(CAMP) at the University of Arizona. 
We are grateful to M. Chandross and Z. Vardeny for critical reading of the
manuscript.

%
\begin{table}
\protect\caption{The $x$ and $y$ components of $1A_g \protect\rightarrow 1B_u$ 
transition dipole moments $\mu_x$ and $\mu_y$ (in A.), as a 
function of  $t_{\perp}$,
for a PDPA oligomer with five double bonds. The results 
for an artificial trans-stilbene
molecule (see text), oriented as in in the inset of Fig. \ref{fig-ssh}, 
are included for comparison.}
  \begin{tabular}{lcccc} 
 \multicolumn{1}{l}{$t_{\perp}$}  &
 \multicolumn{2}{c}{$H_{ee}=0$} & 
\multicolumn{2}{c}{$H_{ee} \neq 0$ (SCI)} \\
       & $\mu_x$  & $\mu_y$ &  $\mu_x$   & $\mu_y$  \\
 0.0   & 2.25  & 0.35  &  1.74   &    0.50   \\
 0.2   & 2.25  & 0.37  &  1.43   &    0.80  \\
 1.4   & 2.17  & 0.98  &  1.39   &    1.08  \\
 ``trans-stilbene''   
                  & 0.34  & 1.08  &  0.19   & 0.59 
    \end{tabular}                      
  \label{tab-edm}    
\end{table}  
\begin{table}  
 \protect\caption{Correlated $2A_g$ and $1B_u$ energies for 
PDPA with four backbone carbons
and butadiene. Sets 1, 2 and 3 refer 
to three different calculations (see text).}
\protect
  \begin{tabular}{cccccccc}
 \multicolumn{1}{c}{System}   & \multicolumn{1}{c}{State} &
 \multicolumn{6}{c}{Energy Gaps (eV)}  \\ 
 & & \multicolumn{3}{c}{$\delta=0.0$} & 
\multicolumn{3}{c}{$\delta=0.07$}\\
  & & Set 1 & Set 2 & Set 3 & Set 1 & Set 2 & Set 3  \\ \hline
PDPA & $2A_g$ & 4.22 & 4.26 & 5.71 & 4.86 & 4.84 & 5.84 \\
     &  $1B_u$ & 4.49 & 4.13 & 4.30 & 4.81 & 4.46 & 4.72 \\
butadiene    & $2A_g$ & 4.66 &         &        &  5.35 &          &         \\
     & $1B_u$ & 5.49 &         &        &  5.80 &          &       
   \end{tabular}                      
  \label{tab-elev1}    
\end{table}  
\end{document}